\documentclass{an}
\usepackage{graphicx}
\usepackage{times}
\Volume{00}              
\Year{0000}              
\Month{00}               
\Pagespan{000}{000}      

\begin{document}
\lhead[\thepage]{T. Becker: Crowded Field 3D Spectroscopy}
\rhead[Astron. Nachr./AN~{\bf XXX} (200X) X]{\thepage}
\headnote{Astron. Nachr./AN {\bf 32X} (200X) X, XXX--XXX}

\title{Crowded Field 3D Spectroscopy}

\author{Thomas Becker\inst{1}, Sergei Fabrika\inst{2}\and Martin M. Roth\inst{1}}
\institute{Astrophysikalisches Institut Potsdam, An der Sternwarte 16, D-14482 Potsdam, Germany\and Special Astrophysical Observatory of the Russian AS, Nizhnij Arkhyz 369167, Russia}
\date{Received {date will be inserted by the editor}; 
accepted {date will be inserted by the editor}} 

\abstract{
The quantitative spectroscopy of stellar objects in complex environments is mainly limited by the ability of separating the object from the background. Standard slit spectroscopy, restricting the field of view to one dimension, is obviously not the proper technique in general. The emerging Integral Field (3D) technique with spatially resolved spectra of a two-dimensional field of view provides a great potential for applying advanced subtraction methods. In this paper an image reconstruction algorithm to separate point sources and a smooth background is applied to 3D data. Several performance tests demonstrate the photometric quality of the method. The algorithm is applied to real 3D observations of a sample Planetary Nebula in M31, whose spectrum is contaminated by the bright and complex galaxy background. The ability of separating sources is also studied in a crowded field in M33.     
\keywords{methods: data analysis - techniques: image processing - techniques: spectroscopic (Integral Field spectroscopy)}
}

\correspondence{tbecker@aip.de}

\maketitle

\section{Introduction}
\vspace{-0.3cm}
3D spectroscopy becomes more and more an important tool not only for spatially extended objects, but also for point sources. This may sound controversial at first. But the more the development of modern telescopes enable access to spectra of individual objects in nearby galaxies, the more the problem of crowding of stars and subtraction of the bright and spatially varying background become limiting factors. Standard slit spectroscopy has a strong disadvantage, as it restricts the field of view to one dimension. Additionaly adverse slit effects are intrinsic limitations for this technique, e.g.:\\
The slit width affects the spectral resolution and the amount of background, which is combined with the object spectrum. On the other hand, when minimizing the slit width, light losses limit the photometric quality of the object spectrum. Atmospheric refraction effects, causing a shift of the object position with wavelength, a wavelength dependant point spread function (PSF), as well as pointing inacurracies are additional factors causing potential light losses differrentially with wavelength. 

The 3D technique can circumvent the instrumental limitations imposed by the slit spectroscopy. It provides the full spectral and spatial information of an observed object and its environment. Interpreting a 3D data cube as a collection of monochromatic images of the identical object rather than a spatially distributed collection of spectra allows the usage of image restoration algorithms to separate the object from its crowded field environment in each wavelength bin. The images are connected in the sense that the shape of the PSF (Fried 1966) as well as the object position (Filippenko 1982) vary as a function of wavelength. Provided the algorithm works with high photometric accuracy, the resulting object intensities of all wavelength bins can be combined to yield a background cleaned object spectrum.\\
The first popular image restoration algorithm was made available by Richardson (1972) and Lucy (1974). With an a priori known PSF, this iterative algorithm sharpens blurred images. However, like similar maximum likelihood algorithms, it does not preserve the photometry, but typically shows large residuals, especially around point sources.\\
In the 1990s much work to improve the RL-algorithm was done at ST-ECF (e.g. Lucy \& Hook 1991, Hook \& Lucy 1993). Lucy (1994) introduced the {\em cplucy} or {\em 2 channel} algorithm which allows the reconstruction of point sources with high photometric accuracy. Instead of a blind deconvolution as performed by the RL-algorithm, the procedure simultaneously models the intensity distribution of a discret number of point sources with {\em known} positions and the background with an assumed smoothness controlled by an additional entropy term. For a detailed description see Lucy (1994), Hook et al. (1994). In the following sections {\em cplucy} and its application to 3D data will be studied. The algorithm will be applied to real 3D data of a point source on a complex but smooth background as well as to a star in a crowded field, surrounded by an emission nebula.

\section{Photometric Performance Tests of {\bf cplucy}}\label{performance}
\begin{figure}[h!]
\vspace{-0.8cm}
\resizebox{\hsize}{!}
{\includegraphics[]{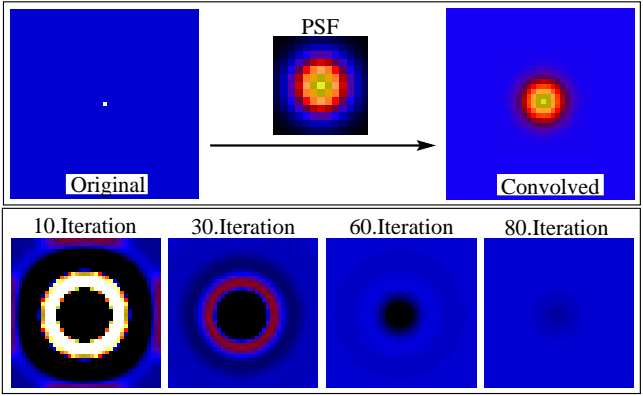}}
\caption{Performance test of the cplucy algorithm. The input is given by a point source on a constant background degraded by a gaussian PSF (upper panel). The lower panel illustrates the residuals in the {\em cplucy} background channel on a $2\%$ scale. For all {\em cplucy}-performance tests described in this paper the relative strength of the entropy term is set to 0.1 (compare Hook et al. 1994). This parameter controls the freedom, with which the background can be adjusted. The standard deviation of the smoothing kernel is set to 10.}
\label{twochan}
\end{figure}
\begin{figure}[h!]
\vspace{-0.3cm}
\resizebox{\hsize}{!}
{\includegraphics[]{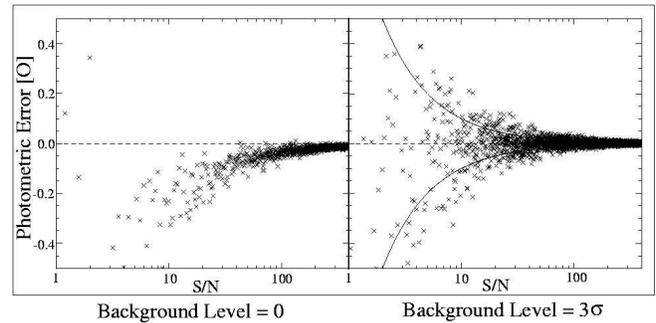}}
\caption{{\em cplucy} performance test as described in Fig. \ref{twochan}. A Poissonian noise with variance $\sigma^2$ is added to the input image. By varying the point source intensity O, we study statistical as well as the systematic errors (see text). The photometric errors are given in units of O. The solid line indicates the statistical noise of the input intensity.}
\label{stocherror}
\end{figure}

 
\vspace{-0.5cm}
The following performance tests study the photometric behaviour of the {\em cplucy} algorithm for a single point source on a constant background. The image is degraded by a known PSF to simulate a seeing-limited exposure . The PSF and the object position are given as input parameters of the algorithm. 
\vspace{-0.4cm}
\begin{quote}
{\bf Test 1:} Study systematic errors and the speed of convergence for various object to background ratios O/B (noise-free simulations).
\end{quote} 
\begin{figure}
\vspace{-0.3cm}
\hspace{0.2cm}
\includegraphics[width=8cm]{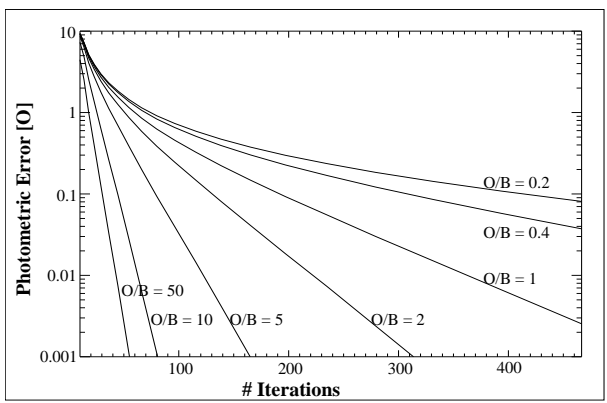}
\caption{{\em cplucy} performance test as described in Fig. \ref{twochan}. Varying the integrated intensity of the point source relative to the backgound level (O/B) demonstrates the speed of convergence. The photometric errors are given in units of the strength of the object signal.}
\label{syserror}
\end{figure}
The initialisation of the algorithm assigns half of the total intensity of the image to the object, half to the background. As shown in Fig. \ref{syserror}, the initial error of the object intensity drops monotonically with growing iteration number and converges to zero. The ratio of object/background (O/B) is in general not constant with wavelength. The speed of convergence however, is a strong function of O/B. In order to avoid different stages of convergence from wavelength to wavelength, a simple procedure was devised: a constant background level or a smooth background model produced by an interpolation method is subtracted before applying {\em cplucy}.
\vspace{-0.15cm}
\begin{quote}
{\bf Test 2:} Study stochastic and systematic errors. A Poissonian noise distribution with a standard deviation of $\sigma$ is added to the image. A previous background subtraction is simulated by setting the background level to 0 ($3\sigma$ in a modified test).  The object intensity is varied from $1$ to $400\sigma$. 
\end{quote}  
\vspace{-0.15cm}
As shown in the left panel of Fig. \ref{stocherror}, the result systematically underestimates the real intensity with decreasing object intensity. The reason is the following: the algorithm sets all negative image values to zero, thus modifying the noise behaviour and raising the mean background level. Adding a constant level $c$ to all the images circumvents this problem. $c$ should be as small as possible to avoid different speeds of convergence for different slices as described above. The right panel of Fig. \ref{stocherror} shows that a value of $c=3\sigma$ where $\sigma$ is the level of background noise is a proper choise. The systematic underestimate vanishes, and the stochastic error is only $15\%$ higher than the noise level of the object input intensity.
\vspace{-0.3cm}
\vspace{-0.8cm}

\section{Photometry of Background-limited Planetary Nebulae in M31}\label{pn}
\begin{figure}[t!]
\resizebox{\hsize}{!}
{\includegraphics[]{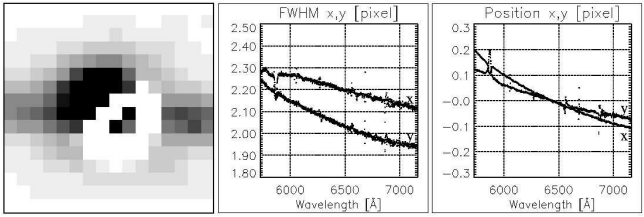}}
\vspace{-0.5cm}
\caption{Moffat fit of a standard star PSF with PMAS. {\bf left:} Fit residuals on a $0-1\%$ scale. {\bf middle:} X,Y PSF width variation with wavelength. {\bf right:} X,Y PSF position variation with wavelength.}
\label{psffit}
\vspace{-0.4cm}
\end{figure}

\begin{figure}[b!]
\vspace{-0.6cm}
\resizebox{\hsize}{!}
{\includegraphics[]{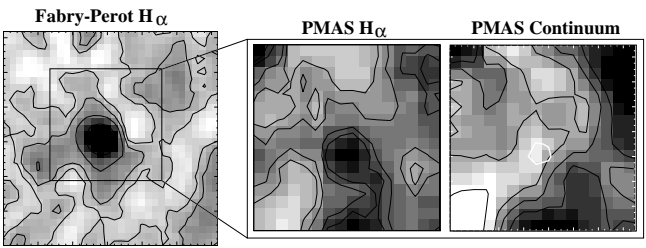}}
\caption{{\bf left:} Cutout of a Fabry-Perot image taken at the Calar Alto 3.5m prime focus. {\bf right:} Monochromatic images from a PMAS 3D exposure. The white contour in the continuum image indicates the position of the PN.}
\label{pn29_mono}
\end{figure}

\begin{figure}
\vspace{-0.2cm}
\resizebox{\hsize}{!}
{\includegraphics[]{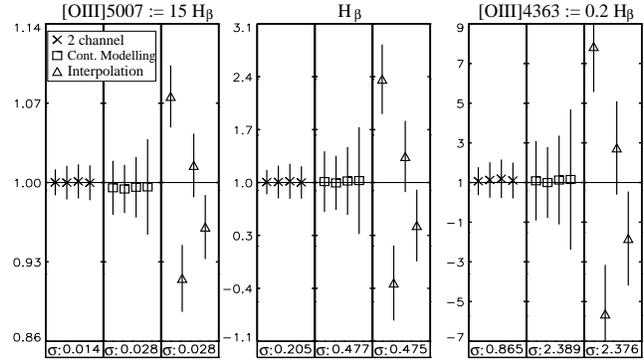}}
\vspace{-.5cm}
\caption{{\em cplucy} performance test with real 3D data taken with PMAS from PN29 (Roth et al. 2003). A model PSF (FWHM=1arcsec) with a strength comparable to three different PN29 emission lines ([OIII]5007,$H_\beta$,[OIII]4363) is put to four different places in the field of view. The test is repeated for all wavelength bins (Test 3), resulting in a mean extracted object intensity plus a standard deviation $\sigma_i$ for each position (four symbols in each subpanel indicating the different positions). The intensities are plotted relative to the input strength. The printed $\sigma$-value at the bottom of each subpanel denotes the mean $\sigma_i$. The {\em cplucy} performance (left subpanels) is compared to two alternative background subtraction methods (see text).} 
\label{pn29_performance}
\vspace{-.5cm}
\end{figure}

\vspace{-0.2cm}
Planetary Nebulae (PN) are excellent candidates to trace the chemical evolution of nearby galaxies (e.g. Dopita 1997). On account of their bright emission in the prominent [OIII]5007 line they are relatively easy to detect. For a detailed analysis of the chemical properties, however, it is necessary to measure lines which are fainter by a factor of $10^{-3}$ to $10^{-4}$. Therefore the subtraction of the bright and complex galaxy background is the major issue limiting the photometric quality of the PN spectra. In several 3D observation campaigns with PMAS\footnote{http://www.aip.de/groups/opti/pmas/OptI\_pmas.html} and MPFS\footnote{http://www.sao.ru/hq/lsfvo/devices/mpfs/mpfs\_main.html} we obtained data from several PNe in the bulge of M31. The results are discussed in Roth et al. (2003). As a major complication it is shown that the diffuse hot gas is present all over the bulge. Its low excitation lines are of comparable or even higher intensities than the corresponding lines of the point source PNe. Using PMAS data of PN29 from the sample of Ciardullo et al. (1989), we demonstrate the application of {\em cplucy} for 3D data of a background-limited point source.\\
A crucial step is the determination of the PSF and the position of the PN. At the distance of M31 (770kpc, Freedman 1990) PNe with a typical diameter or less than 0.4pc appear pointlike for seeing-limited, ground-based observations. This implies that the PSF of the PN is given by the seeing disk which changes slowly as a function of wavelength (Fried 1966). Also the observed PN position changes with wavelength due to the atmospheric refraction effect (Filippenko 1982). Both the position and the shape of the PSF can be measured well for the bright [OIII]5007 emission line and applied to each wavelength bin inside the observed range. Fig. \ref{psffit} demonstrates the quality of how we fit a Moffat function to the 3D data of a standard star. The PSF position and the PSF widths are measured with an inaccuracy of better than 0.002 spaxels. The mean residuals of the fit shown in the left figure are smaller than $1\%$. They originate from a non symmetric PSF of the telescope and crosstalk effects caused by a less than perfect extraction of the spectra. In Becker (2002) the importance of an accurate spectral extraction algorithm is discussed in detail.\\
Fig. \ref{pn29_mono} shows two pseudo-monochromatic maps of PN29, observed with PMAS. The point source is located in a local minimum of the background surface brightness distribution. Conventional background subtraction techniques, in particular with slit spectroscopy, are likely to fail in predicting the correct interpolated background (see Jacoby et al. 1999 and Richer et at. 1999).\\
The continuum emission of the PN is neglegible in comparison with the background. The emission of the PN and of the diffuse gas occur only in some discrete wavelengths. The monochromatic images of the remaining wavelength bins show pure background continuum. The normalized continuum distribution changes smoothly with wavelength, which allows one to model the continuum in each wavelength bin. Based on this consideration another perfomance test of {\em cplucy} was carried out with real 3D data:
\begin{quote}
{\bf Test 3:} A model PSF is inserted at four different positions into a monochromatic continuum image of a PMAS PN29 exposure. {\em cplucy} is run to separate the PSF from the background. Repeating the test for all wavelength bins results in a mean PSF intensity plus a standard deviation $\sigma$ for each of the four positions.  
\end{quote}
Fig. \ref{pn29_performance} demonstrates the results for all object positions, plotting the mean extracted intensities relative to the input intensities. For comparison to the {\em cplucy} results (left subpanels) the extraction results of two alternative extraction methods are plotted in the middle and right subpanels:\\
{\bf Right:} Interpolation of the background from an annulus (32 spaxels in total) centered on the PN (Radius: 2'').\\
{\bf Middle:} Interpolation of the backgound from the same annulus as before, but the continuum distribution was modelled and subtracted beforehand.\\
{\em cplucy} as well as the continuum subtraction method are free from systematic errors contrary to the pure interpolation method. It is important to note that the statistical errors originating mainly from the background noise are much lower for {\em cplucy} compared to the other methods. Because of the spatial variation of the background, only a limited number of sampling points, located near the object, can be used for interpolation methods on account of a larger statistical noise.\\ 
More severly than the continuum, the presence of diffuse gas emission in the vicinity of the PN limits the accuracy of the PN line fluxes (Fig. \ref{pn29_mono}). In more than $70\%$ of PNe observed by Jacoby\&Ciardullo (1999) the presence of systematic errors due to contaminating emission was considered a possibility. Our cplucy analysis show indeed that all the intensities for low excitation lines are systematically lower than those of Jacoby\&Ciardullo (1999), whereas the recombination lines and bright [OIII] lines are in accord within the error bars.\\
The ability to model the background continuum, these targets allow for an optimal preparation of 3D data before applying the {\em cplucy} algorithm: Subtracting the continuum beforehand avoids any substantial differences of speed of convergence. 
\newpage

\section{Crowded field Spectroscopy in M33}\label{lbv}

\vspace{-1.2cm}
\begin{figure}[h!]
{\includegraphics[width=7.5cm]{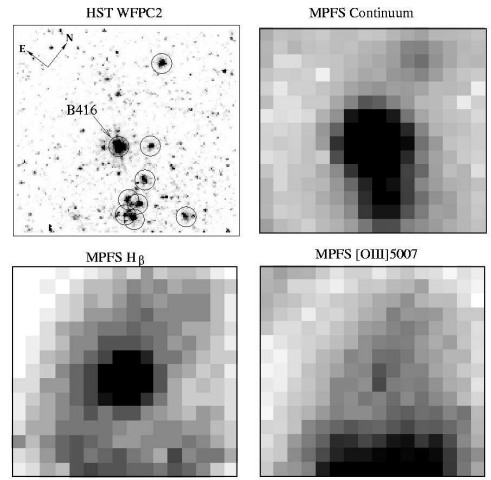}}
\vspace{-0.8cm}
\caption{Monochromatic images of B416 taken from a MPFS 3D exposure. Field of view: 16 arcsec$^2$, spaxel size: 1 arcsec$^2$. {\bf Upper left:} Cutout of a HST WFPC2 exposure matched to the MPFS field of view. This frame is taken for guiding the {\em cplucy} decomposition. Encircled objects are chosen as input point sources.}
\label{b416_mono}
\end{figure}

A more complicated situation of {\em crowding} occurs when several distinct objects are located close to each other such that their PSFs overlap. Fig. \ref{b416_mono} shows an example of such a crowded field in M33, observed with MPFS. The bright star in the center is B416, a LBV candidate, identified as such by Shemmer et al. (2000). The spectroscopic results of these observations are discussed in Fabrika et al. (2003). As seen in the lower maps, B416 is surrounded by a complex nebula with different spatial structures for different emission lines.\\
The upper left image of Fig. \ref{b416_mono} shows a cutout of a HST WFPC2 exposure of the field. This image is used for guiding {\em cplucy}, i.e. to fix the positions of these stars. The circles in the HST image indicate the brightest sources selected as input for {\em cplucy}.\\ 
In the absence of an empirically determined PSF , we applied a gaussian profile. The profile widths were determined in an iterative procedure, applying {\em cplucy} to a continuum image of B416 and minimizing the background residuals.

\begin{figure}[t!]
{\includegraphics[width=7.5cm]{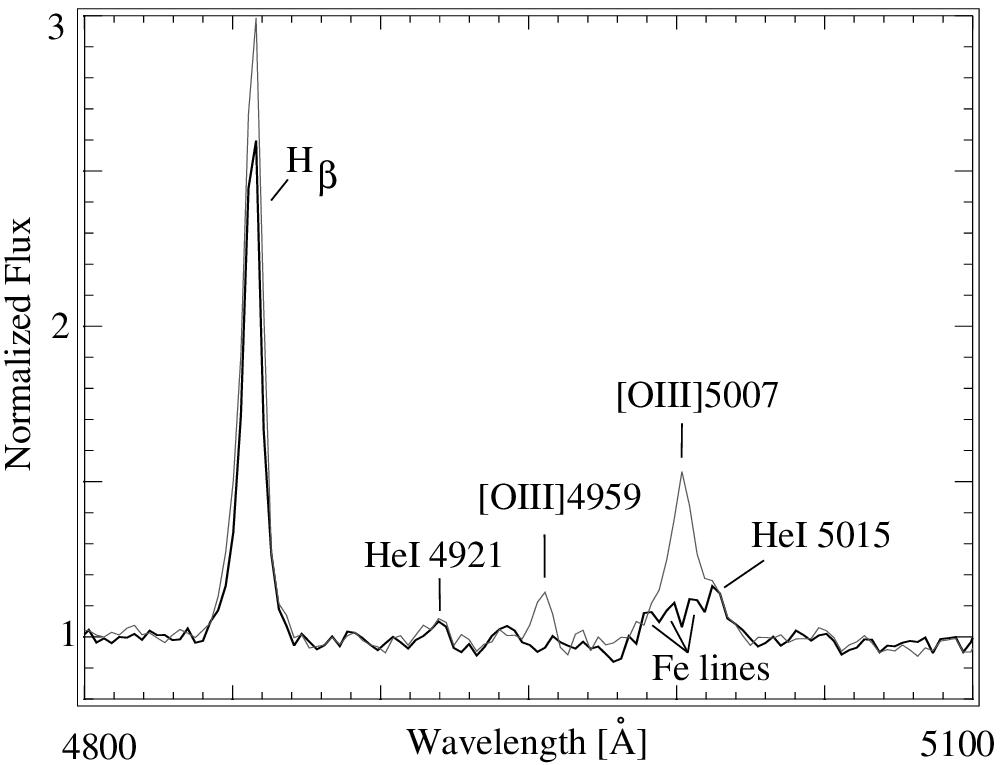}}
\caption{Part of the B416 spectrum after applying cplucy to the MPFS data. The thin line simulates a slit spectrum. The background was subtracted by interpolating the ends of the slit to the star position. Note the remaining emission originating from the nebula.}
\label{b416_plot}
\end{figure}

With the given coarse sampling and low spatial resolution (seeing 3 arcsec FWHM), it was impossible to deconvolve the faint individual stars. The important point, however, was our ability to treat the nebular emission as a smoothly varying component of the background channel, thus separating - at least - the bright star from the nebula. From another observation with long slit spectroscopy, the star was known to exhibit weak [O III] 5007 and 4959 emission, which was speculatively attributed to an unresolved, small nebula, which would be different from the larger shell. This spectrum has been simulated from the 3D data in Fig \ref{b416_plot}. After applying {\em cplucy}, the spectrum of B416 looks quite different at these wavelengths, and the [O III] emission has disappeared. Instead, a blend of faint Fe I lines and He I 5015 is seen, whose origin is very likely B416.

\vspace{-0.3cm}
\section{Summary}
\vspace{-0.3cm}
We have demonstrated that the cplucy algorithm can be used to improve the image quality of 3D data. Applied to datacubes of a faint, background limited planetary nebula in M31, it is shown that systematic errors of emission line intensities, which seem to be unavoidable with the limited geometry of slit spectroscopy, can be eliminated by using the knowledge of the PSF. The absence of a measurable continuum of the PN is a favourable case, since it allows one to model the background continuum in between the PN emission lines. Subtracting the background model prior to iterating cplucy helps to make sure that the algorithm is converging with the same speed in each wavelength bin. With an example of the luminous star B416 in M33, located in a gaseous emission region, we have also shown that the cplucy  method is even useful in cases where the image quality does not seem to warrant any deconvolution effort, helping nevertheless to disentangle the stellar PSF from the gaseous background component.

\end{document}